\documentclass[sn-mathphys]{sn-jnl}% Math and Physical Sciences Reference Style
\usepackage{amssymb,amsmath}%
\jyear{2022}%

\theoremstyle{thmstyleone}%
\theoremstyle{thmstyletwo}%

\theoremstyle{thmstylethree}%

\newcommand{\Thot}{T_{1}^*}
\newcommand{\Tcold}{T_{2}^*}
\newcommand{\Tref}{T_{0}^*}
\newcommand{\gr}{\mbox{Gr}}

\newcommand{\cs}{\mathcal{C}}
\newcommand{\si}{\mathcal{S}}

\theoremstyle{definition}
\newtheorem*{notation}{\bf Notations}

\begin{document}
\title[\empty]{Longitudinal and transverse modes of temperature modulated inclined layer convection}


\author*[1]{\fnm{Jitender} \sur{Singh}}\email{sonumaths@gmail.com}
\affil*[1]{\orgdiv{Department of Mathematics}, \orgname{Guru Nanak Dev University}, \orgaddress{\street{}, \city{Amritsar}, \postcode{143005}, \state{Punjab}, \country{India}}}
%\affil[2]{\orgdiv{Department}, \orgname{Organization}, \orgaddress{\street{Street}, \city{City}, \postcode{10587}, \state{State}, \country{Country}}}
%\affil[3]{\orgdiv{Department}, \orgname{Organization}, \orgaddress{\street{Street}, \city{City}, \postcode{610101}, \state{State}, \country{Country}}}

\abstract{A parametric instability of an incompressible, viscous, and Boussinesq fluid layer bounded between two parallel planes is investigated numerically. The layer is assumed to be inclined at an angle with horizontal. The planes bounding the layer are subjected to a time-periodic heating. Above a threshold value, the temperature gradient across the layer leads to an instability of an initially quiescent state or a parallel flow, depending upon the angle of inclination. The Floquet analysis of the underlying system reveals that under modulation, the instability sets in as a convective roll pattern executing  harmonic or subharmonic oscillations, depending upon the modulation, the angle of inclination, and Prandtl number of the fluid.  Under modulation, the value of the angle of inclination for  the codimension-2 point is found to be a nonconstant function of the amplitude and the frequency of modulation. Further, the instability response in the fluid layer as a longitudinal mode is always harmonic whereas the instability response as a transverse mode is harmonic, or subharmonic, or bicritical depending upon the modulation. The temperature modulation offers a good control of time-periodic heat and mass transfer in the inclined layer convection.}

%%================================%%
%% Sample for structured abstract %%
%%================================%%




\maketitle

\section{Introduction}
Inclined layer convection (hereafter referred to as ILC) is the hydrodynamic problem of instability in a differentially heated layer of a viscous, incompressible, and Newtonian fluid bounded between two parallel planes inclined at some angle with the horizontal  \cite{Tritton1963,kierkus1968analysis,sparrow1969longitudinal,Llyod1970sparrow,clever1973finite,hollands1976free,clever1977instabilities,hideo1984}. The onset of ILC is generally of two types:

(i) the buoyancy dominated instability associated with the well studied \emph{Rayleigh-B\'enard convection} (hereafter referred to as RBC) in an initially quiescent horizontal fluid layer \cite{chndra,drazin,koschmieder,bodenschatz2000}, and

(ii) the one with the dynamic type of the well understood shear dominated vertical layer natural convection (hereafter referred to as VLC)  \cite{ref0,ref1,ref2,korpelaetal,bergholz1978,suslov}.

As the angle of inclination of the fluid layer with the horizontal varies from zero, the mechanism of the onset of ILC is of RBC type  upto an inclination of about $71^\circ$ beyond which the instability mechanism is of the VLC type \cite{hart1971stability}.
%In a remarkable experimental work,  Tritton \cite{Tritton1963} explored on detection of the flow transition  from laminar to turbulent in ILC.  Kierkus \cite{kierkus1968analysis} performed experiments on ILC in measuring velocity of the flow field.

In \cite{sparrow1969longitudinal}, Sparrow and Husar carried out experimental studies on ILC and found that for the considered range of inclinations, the onset of ILC in air is in the form of longitudinal rolls. Further, in another experimental work, Lloyd and Sparrow \cite{Llyod1970sparrow}  determined the dependence of Rayleigh number for the onset of ILC  on the inclination of the fluid layer. It was found that for the inclination of the fluid layer with the horizontal upto $73^\circ$, the instability is characterized by longitudinal vortices, while for the inclination beyond $76^\circ$, the instability is characterized by the transverse vortices, where for the inclinations between $73^\circ$-$76^\circ$ a  transition occurs between the two modes of the instability.

These results were further confirmed by  experimental and   detailed theoretical work of Hart \cite{hart1971stability} on ILC with an excellent  correlation between theory and experiment.

Clever and Busse \cite{clever1977instabilities} performed a numerical study on the stability analysis of longitudinal rolls in ILC and found a possibility of three types of transitions from the longitudinal rolls to the three dimensional form of convection, depending upon the inclination of the fluid layer.

Hideo \cite{hideo1984} studied experimentally the flow and heat transfer behavior of  ILC in a finite box of large aspect ratio revealing a good agreement with the past theoretical studies.

Daniels  et al. \cite{PhysRevLett.84.5320} performed experiments on ILC for a fluid of Prandtl number 1 and found many new interesting nonlinear chaotic states. These findings have been found to be in correlation with the recent numerical work of Subramanian et al. \cite{Subramanian2016}, Reetz and Schneider \cite{Reetz2020A22}, Reetz et al. \cite{Reetz2020A23}, and Tuckermann \cite{Tuckerman2020} on ILC, where a variety of spatio-temporal patterns have been found to be exhibited by ILC for the fluids having  Prandtl number near unity.

A rich variety of patterns exhibited by ILC as evident from the aforementioned experimental and theoretical research witnesses the technological importance of ILC for further scientific investigations. Moreover, ILC has served as a model problem for practical utility from micro scale to mega scale \cite{Arora2020} in a number of heat transfer, material processing, and other industrial applications.

The heat and mass transfer characteristics of thermal convection in ILC can be controlled by an external time-periodic forcing of the bounding planes. In view of this, the parametric excitation of Faraday instability \cite{edwardfauve1994,bessonetal1996,Arbell2002,Batson2015}, RBC under time-periodic temperature modulation \cite{bodenschatz2000,singh2008,Somorodin2009,singh2009,singh2011,singh2015,singh2016,Kaur2016} or gravity modulation \cite{Farooq1996,volmermuller1997,rbajaj1997,chenchen1999}, and time-periodic temperature or gravity modulation of VLC \cite{Baxi1974,SB2018,SB2020} are known to execute harmonic, or subharmonic, or quasi-periodic oscillations at the onset of the instability, depending upon the modulation and Prandtl number of the fluid. The literature is vast on the past researches on these themes, and the reader is referred to go through the excellent reviews in \cite{davis1976,mileshenderson1999,bodenschatz2000}. Each of these time-periodic modulation types can lead to advancement  or delay of the onset of convection in different parametric ranges of the frequency and the amplitude of modulation.

Using Floquet theory Singh and Bajaj \cite{SB2018} have performed a linear instability analysis of VLC under time-periodic modulation of the temperatures of the vertical planes bounding the layer for the fluids with Prandtl number $<12.5$. Besides the destabilizing and stabilizing effects of modulation parameters on the onset of the instability, they observed that the onset of the instability is either a harmonic, or a subharmonic mode  where the mode switching always occurs through an intermediate bicritical state. For a particular combination of amplitude and frequency of forcing, the bicritical state may correspond to coexistence of a pair of either purely harmonic modes, or purely subharmonic modes, or one harmonic and one subharmonic mode. Further, for fluids with Prandtl number $<12.5$, Singh et al. \cite{SB2020} have found a much wider parameter space for observing bicritical states exhibited by the temperature modulated VLC under two-frequency forcing of the temperatures of the vertical planes bounding the layer.

The practical utility of the temperature modulated ILC (hereafter referred to as TMILC) offers a good control over time-periodic heat and mass transfer. In fact  TMILC has a wide range of applications in  the processes  where rapid heat and mass transfer is required (e.g. designing heat sinks for cooling of electronic devices, heat exchangers in nuclear reactors, air conditioning, microwaves,  etc.)  In this view, only the two particular configurations of TMILC, that is, the temperature modulated RBC and the temperature modulated VLC have been investigated in detail in the recent past \cite{SB2018,SB2020}, which restrict its application in the case of a tilted configuration of the layer. This serves as a motivation for the present work for investigating the hydrodynamic behavior of TMILC for all possible configurations. So, in the present paper, within the framework of linear instability theory, we examine TMILC  for the fluids with Prandtl number $<12.5$ under single frequency time-periodic excitation of the temperatures of the two planes bounding the fluid-layer. The Floquet theory is utilized for the purpose.

The paper is organized as follows. The problem and the basic state are discussed in Sec.~\ref{sec:1}, where a linear instability analysis of the basic state is carried out. In Sec.~\ref{sec:2},  Floquet analysis of the underlying linear system is performed and the problem is reduced to an equivalent generalized eigenvalue problem for the control parameter. Most of the numerical results are discussed in Sec.~\ref{sec:3} for the Prandtl number of air. The effect of Prandtl number on the angle of inclination corresponding to the codimension-2 point in TMILC is discussed separately in Sec.~\ref{sec:4}. The  conclusions  are presented in Sec.~\ref{sec:5}.
\section{Mathematical Formulation}\label{sec:1}
\begin{figure}[h!!!]
\vspace{-1cm}
\centering
\includegraphics[width=0.6\textwidth,angle=-45]{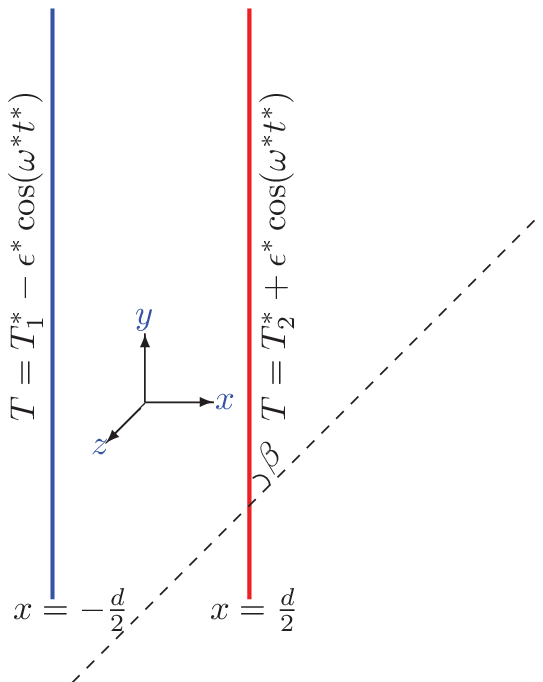}
\vspace{-2cm}
\caption{Geometry of TMILC. }\label{F0}
\end{figure}

We consider a layer of thickness $d>0$ of a viscous, incompressible,  and Newtonian fluid between  two rigid parallel planes inclined at an angle $\beta$ degrees with respect to the horizontal. Affixing the coordinate system with the fluid layer as shown in Fig.~\ref{F0}, the planes bounding the layer are $x=-d/2$ and $x=d/2$, which  are maintained at temperatures $\Thot-\epsilon^*\cos(\omega^*t^*)$ and $\Tcold+\epsilon^*\cos(\omega^*t^*)$, respectively with the base frequency $\omega^*>0$ and amplitude of modulation $\epsilon^*\geq 0$, such that $\Thot>\Tcold\geq 0$.

We introduce the scales to measure length, time, velocity, pressure, and temperature as $d$, $1/\omega^*,$ %$d^2/\kappa$,
$\kappa/d$, $\rho_0\kappa^2/d^2$, and  $\Delta T^*=\Thot-\Tcold$, respectively where $\kappa$ is the thermal diffusivity of the fluid and $\rho_0$ is the density of the fluid at temperature $\Tref=(\Thot+\Tcold)/2$.  Using these scales in laws of conservation of mass, momentum, and energy, we obtain the Grashof number $\gr=\displaystyle {\alpha d^3 \rho_0^2 g\Delta T^*}/{\eta^2}$, the Prandtl number $\sigma={\eta}/{(\rho_0\kappa)}$, the amplitude of modulation $\epsilon={\epsilon^*}/{\Delta T^*}$, and the basic frequency of modulation $\omega={d^2\omega^*}/{\kappa}$ as the four dimensionless parameters governing the flow, where $g$ and $\eta$ denote the gravitational acceleration and the dynamic viscosity of fluid, respectively. Also, $\alpha=\frac{1}{\rho_0}\frac{\partial \rho}{\partial T^*}$, where the fluid density $\rho$ at temperature $T^*$ is given by the
equation of state as follows:
\begin{equation}
    \rho=\rho_0\{1-\alpha(T^*-\Tref)\}.
\end{equation}
In view of the aforementioned scaling, we
further scale the temperature field within the layer so that
\begin{equation}
    T = {(T^*-\Tref)}/{\Delta T^*}
\end{equation}
is the dimensionless form of the temperature. So, mathematically the working domain of the fluid layer is the set $\mathcal{D} = (-\frac{1}{2},\frac{1}{2})\times\Bbb{R}\times \Bbb{R}$.
The basic state in dimensionless form consists of the basic velocity field $(0,\sigma \gr V_e(x,t)\sin\beta,0)$ obeying zero net flux  $\int_{-{1}/{2}}^{{1}/{2}}V_e(x,t)dx=0$,  the basic temperature field $T=T_e(x,t)$, and the basic pressure $P=P_e(x,t)$ associated with the fluid layer which are given as follows:
%\begin{subequations}\label{e2}
   \begin{equation} \label{e2a}
V_e(x,t)=\frac{x}{6}\Bigl(x^2-\frac{1}{4}\Bigr)+\epsilon{\sigma} \mathfrak{Re} \Bigl\{\frac{f_1(x,\omega)-f_{\sigma}(x,\omega)}{(1-\sigma)\iota\omega}
 e^{\iota t}\Bigr\},
 \end{equation}
\begin{equation}\label{e2b}
T_e(x,t)=-x+\epsilon \mathfrak{Re}\bigl\{f_1(x,\omega)e^{\iota t}\bigr\},
\end{equation}
\begin{equation}\label{e2c}
P_e(x,t)=P_e(-1/2)-\frac{d^3g}{\kappa^2}(x\cos \beta+y\sin\beta)-\sigma^2\gr\cos\beta\int_{-1/2}^x T_e(s,t)ds,
\end{equation}
%\end{subequations}
where  $\mathfrak{Re}[\cdot]$ denotes the real part of $[\cdot]$, and
\begin{equation}
f_\sigma(x,\omega)={\sinh\{\sqrt{{\iota\omega}/{\sigma}}x\}}/{\sinh\{(1/2)\sqrt{{\iota\omega}/{\sigma}}\}},
\end{equation}
where we have  ${f_1}={f_\sigma\mid}_{\sigma=1}$.

Considering small perturbations in \eqref{e2a}-\eqref{e2c} so that the perturbed velocity $(0,\sigma \gr V_e,0)+(u,v,w)$, the perturbed temperature  $T_e+\theta$, and the perturbed pressure $P_e+P$ satisfy the governing equations subject to the no-slip conditions at the rigid walls; the small perturbations $u,v,w,\theta,P$ after retaining linear terms and possibly eliminating $v$, $w$ and the pressure term $P$ from the governing equations lead to the following PDEs
%\begin{subequations}
\begin{eqnarray}\label{e3}
   \nonumber\frac{\omega}{\sigma} \frac{\partial \nabla^2u}{\partial t}+\gr\sin\beta\Bigl(V_e\nabla^2-\frac{\partial^2V_e}{\partial x^2}\Bigr)\frac{\partial u}{\partial y}&=&\nabla^4u-\sigma \gr\sin\beta\frac{\partial^2 \theta}{\partial y\partial x}\\&+&\sigma\gr\cos\beta\Bigl(\frac{\partial^2\theta}{\partial y^2}+\frac{\partial^2\theta}{\partial z^2}\Bigr),
\end{eqnarray}
\begin{eqnarray}\label{e4}
\omega\frac{\partial \theta}{\partial t}+u\frac{\partial T_e}{\partial x}+\sigma \gr\sin\beta V_e\frac{\partial \theta}{\partial y}=\nabla^2 \theta,
\end{eqnarray}
along with the boundary conditions given by
\newpage
\begin{eqnarray}\label{e5}
\Bigl(u, \frac{\partial u}{\partial x}, \theta\Bigr){\mid}_{x=\pm\frac{1}{2}}=(0,0,0).
\end{eqnarray}
%\end{subequations}
\section{Method of solution}\label{sec:2}
In view of the fact that the perturbations remain bounded on $\mathcal{D}$ and periodic in  $y$ and $z$ and $t$, we expand $u$ and $\theta$ in the following appropriate Fourier-Floquet form \cite{chndra,kumar1994} \begin{subequations}
\begin{eqnarray}\label{e6a}
  u &=& \sum_{q=-L}^L\sum_{\ell=1}^{N}\{a_{\ell q}\Phi_\ell(x)+\iota b_{\ell q}\Psi_{\ell}(x)\}e^{\iota(s+q)t}e^{\iota\mathbf{k}\cdot\mathbf{z}}\\
  \label{e6b}\theta &=& \sum_{q=-L}^L\sum_{\ell=1}^{N}\{\iota c_{\ell q}\si_{2\ell}(x)+d_{\ell q}\cs_{2\ell-1}(x)\}e^{\iota(s+q)t}e^{\iota\mathbf{k}\cdot\mathbf{z}}
\end{eqnarray}
\end{subequations}
where $\mathbf{z}=(0,y,z)$ and $\mathbf{k}=(0,k\cos \gamma,k\sin \gamma)$ for $0\leq \gamma\leq \frac{\pi}{2}$
 is the wave vector of perturbations with the wave number $k>0$, and the number of Galerkin terms $N$ in the above expansions is chosen large enough to meet numerical convergence. For each $\ell$, the functions $\Phi_\ell$, $\Psi_\ell$, $\si_\ell$, and $\cs_\ell$ are known as Chandrasekhar functions \cite{chndra} and are defined as
\begin{eqnarray*}
\Phi_\ell(x)=\displaystyle \frac{\cosh{\lambda_\ell
x}}{\cosh{\frac{\lambda_\ell}{2}}}-\frac{\cos{\lambda_\ell
x}}{\cos{\frac{\lambda_\ell}{2}}};~
\Psi_\ell(x)=\displaystyle \frac{\sinh{\mu_\ell
x}}{\sinh{\frac{\mu_\ell}{2}}}-\frac{\sin{\mu_\ell
x}}{\sin{\frac{\mu_\ell}{2}}},
\end{eqnarray*}
\begin{eqnarray*}
\si_\ell(x)=\sin \{\ell\pi x\};~\cs_{\ell}(x)=\cos\{\ell\pi x\},
\end{eqnarray*}
where $\lambda_\ell$ and $\mu_\ell$ satisfy the following:
\begin{eqnarray*}
\tan\frac{\lambda_\ell}{2}+\tanh\frac{\lambda_\ell}{2}=0;~\cot\frac{\mu_\ell}{2}-\coth\frac{\mu_\ell}{2}=0.
\end{eqnarray*}
Finally, $s=0$ and $s=1/2$ correspond to the harmonic and the subharmonic responses of the perturbations, respectively.

Substituting the truncated expansions for $u$ and $\theta$ from \eqref{e6a}-\eqref{e6b} into \eqref{e3}-\eqref{e4} and performing Galerkin operations by taking unity as the weight function, we obtain the following:
\begin{equation}\label{e170}
\begin{split}
\{\mathbf{L}_q-\sigma\gr(\mathbf{U_1}\cos\beta+\mathbf{U_2}\sin\beta)\}\zeta_q-\epsilon \sin\beta (\mathbf{V}+\sigma\gr\mathbf{W})\zeta_{q-1}\\-\epsilon \sin\beta (\bar{\mathbf{V}}+\sigma\gr\bar{\mathbf{W}})\zeta_{q+1} =0,
\end{split}
\end{equation}
where $\mathbf{L}_q$, $\mathbf{U}_1$, $\mathbf{U}_2$, $\mathbf{V}$, and $\mathbf{W}$, are square matrices defined in the Appendix \ref{appen1}, and for each $q$,
\begin{eqnarray*}
\zeta_q=(a_{1q},\ldots,a_{Nq}, b_{1q},\ldots,b_{Nq},c_{1q}\ldots,c_{Nq},d_{1q},\ldots,d_{Nnq})^{\rm t}
\end{eqnarray*}
is the $4N\times 1$ matrix of unknowns. The system  \eqref{e170} leads to the following generalized eigenvalue problem.
\begin{equation}\label{e171}
\mathbf{L}\zeta =\sigma \gr (\mathbf{Z}_1\cos\beta+\mathbf{Z}_2\sin\beta)\zeta,
\end{equation}
where $\mathbf{L}$, $\mathbf{Z}_1$, and $\mathbf{Z}_2$ are block matrices given by
\begin{equation*}\label{e172}
\mathbf{L}=\left(
  \begin{array}{cccccc}
    \ddots  &\vdots & \vdots & \vdots&\vdots\\
%    \cdots &-\mathbf{L}_{-2}&\epsilon\overline{\mathbf{V}}e^{-\iota\phi}\sin\beta & \mathbf{O} & \mathbf{O} & \mathbf{O} &  \cdots\\
    \cdots & -\mathbf{L}_{-1}&\epsilon\overline{\mathbf{V}}\sin\beta   & \mathbf{O} &\cdots\\
    \cdots & \epsilon{\mathbf{V}}\sin\beta & -\mathbf{L}_{0} & \epsilon\overline{\mathbf{V}}\sin\beta  & \cdots\\
    \cdots &\mathbf{O}  & \epsilon {\mathbf{V}}\sin\beta  & -\mathbf{L}_{1} &\cdots \\
 %   \cdots &\mathbf{O}&\mathbf{O} & \mathbf{O} & \epsilon {\mathbf{V}}e^{\iota\phi}\sin\beta & -\mathbf{L}_{2}&\cdots\\
    \vdots  &\vdots & \vdots & \vdots & \ddots
  \end{array}
\right);~\zeta=\left(
           \begin{array}{l}
              \vdots\\
             %\zeta_{-2} \\
             \zeta_{-1} \\
             \zeta_0 \\
             \zeta_1 \\
             %\zeta_2 \\
             \vdots
           \end{array}
         \right),
\end{equation*}
\begin{equation*}\label{h}
  \mathbf{Z_1}=\left(
  \begin{array}{cccccc}
    \ddots & \vdots & \vdots &  \vdots &\vdots\\
  %  \cdots  &\mathbf{O}& \mathbf{O} & \mathbf{O} &   \cdots\\
    \cdots  &\mathbf{U}_1 & \mathbf{O}& \mathbf{O}   &\cdots\\
    \cdots  &\mathbf{O}& \mathbf{U}_1 & \mathbf{O}  &\cdots\\
    \cdots  &\mathbf{O} & \mathbf{O}& \mathbf{U}_1 &\cdots \\
   % \cdots  &\mathbf{O} & \mathbf{O} & \mathbf{O}& \cdots\\
    \vdots  &\vdots & \vdots & \vdots  &\ddots
  \end{array}
\right);~
%\end{equation*}
%\begin{equation*}\label{e173}
\mathbf{Z_2}=\left(
  \begin{array}{cccccc}
    \ddots  &\vdots & \vdots &  \vdots &\vdots\\
    \cdots &\mathbf{U}_2 & \epsilon \bar{\mathbf{W}}& \mathbf{O} & \cdots\\
    \cdots &\epsilon {\mathbf{W}} & \mathbf{U}_2 & \epsilon \bar{\mathbf{W}}&\cdots\\
    \cdots &\mathbf{O} & \epsilon {\mathbf{W}} & \mathbf{U}_2 & \cdots \\
    \vdots  &\vdots & \vdots & \vdots & \ddots
  \end{array}
\right).
\end{equation*}
We solve the generalized eigenvalue problem \eqref{e171} numerically on computer in order to obtain $\gr^{-1}$ as the real eigenvalue after feeding  a trial value of $k$ in the interval $[0,8]$ for fixed values of the other parameters. The procedure is repeated for the other values of $k$.
The critical Grashof number for the onset of the instability is then computed using the following formula
\begin{eqnarray}\label{ralc}
\gr_c=\min_{s}\inf_{k,\gamma}\gr(\sigma,\epsilon,\omega,\beta,s,k).
\end{eqnarray}
The critical wave number $k_c$ is the value of $k$ corresponding to $\gr_c$.

For numerical computations, we have taken $0\leq \epsilon \leq 0.5$, $0^\circ\leq  \beta <180^\circ$,  and $0<\omega\leq 200$. In this work, most of the numerical calculations have been performed for $\sigma=0.71$ (air). Dependence of the onset and nature of TMILC on $\sigma$ is discussed separately.
\begin{notation}
Throughout, we shall denote by  $k_c^{H}$ and $k_c^{S}$,  the critical wave numbers corresponding to harmonic and  subharmonic types of modes, respectively in TMILC. Also, it has been found that the onset of the instability of ILC is either a longitudinal mode ($\gamma=90^\circ$) or a transverse mode ($\gamma=0^\circ$). We shall use the notation $k_c^L$ and $k_c^T$ for the wave numbers corresponding to longitudinal and transverse modes of unmodulated ILC. Under modulation, we shall denote by $k_c^{LH}$, the longitudinal harmonic mode;  $k_c^{TH}$ or $k_c^{TS}$ to denote the correspondingly transverse harmonic or transverse subharmonic mode of TMILC. Finally, the symbol $\beta_c$ will denote the value of $\beta$ corresponding to the codimension-2 point in TMILC. Thus, for a given set of other fixed parametric values, the preferred mode of the onset of TMILC is longitudinal for $\beta< \beta_c$  and transverse for $\beta>\beta_c$.
\end{notation}
\section{Numerical results and discussion}\label{sec:3}
The equation \eqref{ralc} has been solved numerically  with the help of {\tt MATLAB} programming in order to obtain the critical value of the control parameter for a given set of the other dimensionless parameters. To verify correctness of the numerical code, we have compared the numerical results obtained using the present numerical scheme with those of Hart \cite{hart1971stability} in  Table \ref{t1}  for the unmodulated ILC for the Prandtl number of air, that is, $\sigma=0.71$.
\begin{table}[h!!!]\label{t1}
\centering
\caption{Comparison of present numerical results with those obtained in Hart \cite{hart1971stability} for the transverse mode of unmodulated ILC.}
\begin{tabular}{llcc}
  \hline
 $\beta$ & $\sigma$ & $\gr$ &  $\gr$  \\
 && (Present) & Hart \cite{hart1971stability} \\
 \hline\hline
$0^\circ$  & 6.7  & 254.8898& 254.9253\\%0.0139\%
$12^\circ$ & 6.7  & 278.3455& 275.9701\\%0.853\%
$24^\circ$ & 6.7  & 509.8482& 500.0000\\%1.931\%
$12^\circ$ & 0.71 & 2587.028& 2569.014\\%0.6963\%
%$24^\circ$ & 0.71 & 3540.451& 3798.591\\%7.291\%
  \hline
  \hline
\end{tabular}
\end{table}
Clearly, the values of $\gr$ as obtained using the present numerical method are in good agreement with the corresponding values given in Hart \cite{hart1971stability}. The partial deviation in the numerical values of $\gr$ might be due to the difference in the numerical methods used.
Moreover, for $\epsilon=0$ and $\sigma=1.07$, the angle of inclination  corresponding to the codimension-2 point   has been found to occur for
\begin{eqnarray*}
(k_c^L,k_c^T,\gr_c)=(3.117875,2.830625,7526.231081),
\end{eqnarray*}
where $\beta_c=77.7567^\circ$ for $N=10$, which is very close to the corresponding value $\beta_c=77.7560^\circ$ as obtained recently in \cite{Subramanian2016}.

Finally, for $\epsilon>0$, we have  the following numerical values for $\beta=90^\circ$
 \begin{eqnarray*}
(\sigma,\epsilon,\omega,k_c^H,k_c^{S},\gr_c)=(0.71,0.3406,5,3.033,2.648,8561.187812)
\end{eqnarray*}
for $N=10$ and $L=30$ which are in close agreement with the corresponding values obtained by Singh and bajaj \cite{SB2018}. These observations verify the correctness of our numerical code.

After several numerical experiments, we have found that the present numerical scheme converges for $N\geq 10$ and $L\geq 20$, where for $\beta>150^\circ$ larger values of  $N\geq 15$ and $L\geq 30$ are required to meet the numerical convergence.  We have taken care of fixing $N$ and $L$ accordingly.
\subsection{Marginal instability curves}
\begin{figure}[h!!!]
\centering
\includegraphics[width=\textwidth]{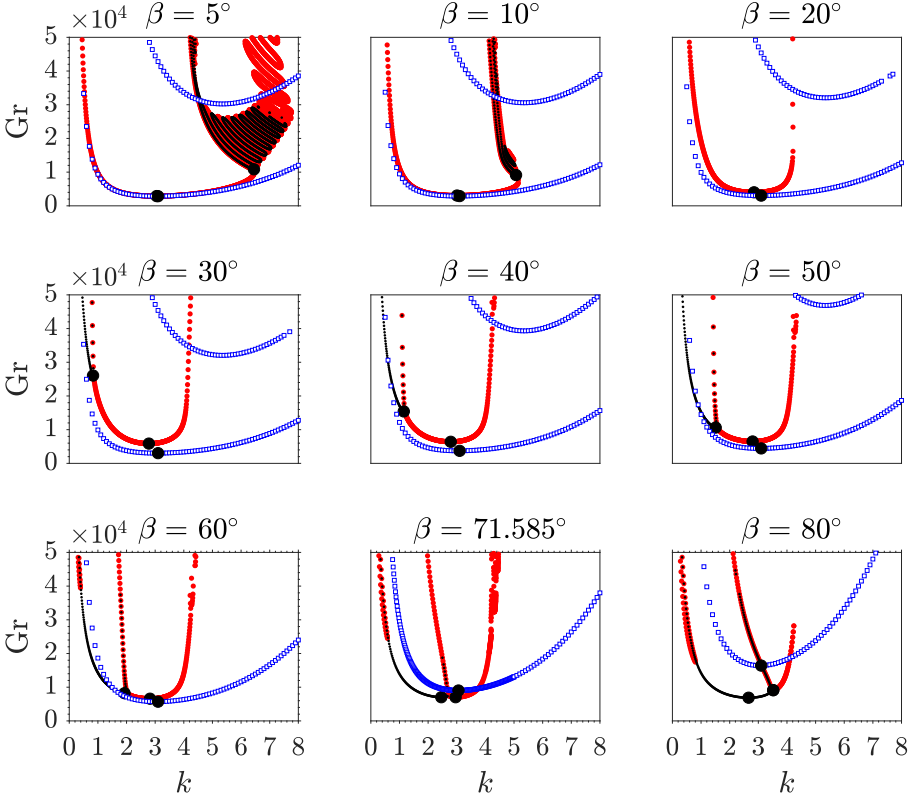}%
 \caption{The neutral instability curves in $(k,\gr)$-plane for $\sigma=0.71$, $\epsilon=0.5$, and $\omega=5$. The  points marked {\color{blue}$\square$} in blue correspond to the longitudinal harmonic mode. The larger (red) and smaller (black) points correspond to the transverse harmonic and subharmonic instability responses respectively. The point marked $\bullet$ in each case denotes the minimum for the particular type of instability mode.}\label{F1a}
\end{figure}

\begin{figure}[h!!!]
\centering
 \includegraphics[width=0.9\textwidth]{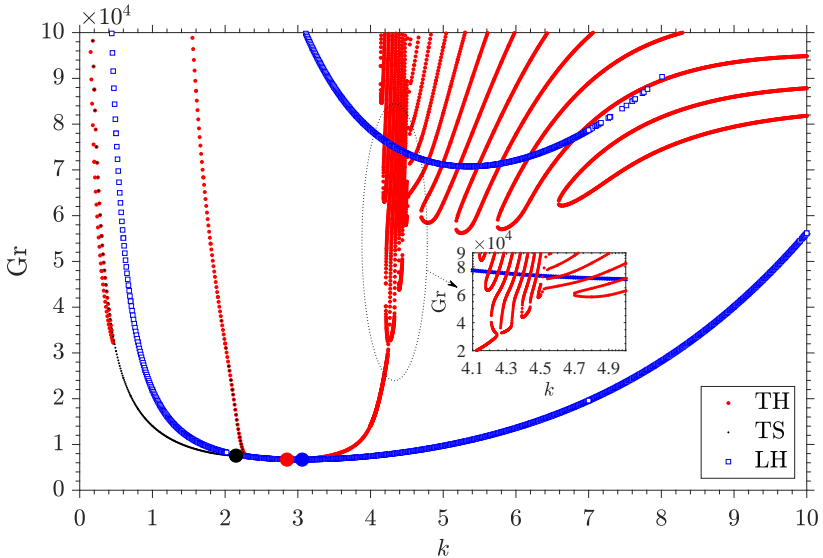}
 \caption{Neutral instability curve in $(k,\gr)$-plane  for $\beta=64.806^\circ$. The other fixed parametric values are same as in Fig.~\ref{F1a}. Each of $\bullet$, {\color{red}$\bullet$}, and {\color{blue}$\bullet$} denote the minimum for TS, TH, and LH responses respectively.}\label{F1b}
\end{figure}

Figure \ref{F1a} shows the marginal curves in $(k,\gr)$-plane for $\sigma=0.71$, $\epsilon=0.5$, and $\omega=5$.
Each subfigure in Fig.~\ref{F1a} has been obtained for one value of $\beta$.  In each of the subfigures, the  points marked {\color{blue}$\square$} in blue correspond to the longitudinal harmonic(LH) mode. The larger (red) and smaller (black) points correspond to the transverse harmonic(TH) and transverse subharmonic(TS) instability responses respectively. The point marked $\bullet$ in a particular marginal curve corresponds to the minimum for the particular type of instability mode. The marginal curves for $\beta=0^\circ$ consists only of a single harmonic branch \cite{SB2018}. However, even for the small value of $\beta=5^\circ$, the marginal instability curves  for the transverse mode consist of alternate harmonic and subharmonic tongues, where a comparatively wide harmonic instability tongue is followed by narrow pattern of alternate subharmonic and harmonic closed curves. The basic state is unstable for the points within each such closed curve and stable outside it.  The lowest value of $\gr$ for the transverse mode occurs on the wider harmonic branch of the marginal curves. On the other hand, the marginal curve for the longitudinal mode consists of a single wide harmonic tongue on which $\gr_c$ is attained and no subharmonic response is found to occur. For $\beta=5^\circ$, the critical value of the control parameter for the temperature modulated ILC corresponds to a longitudinal-harmonic(LH) mode. The pattern of closed TH and TS curves appears for up to $\beta=10^\circ$, where the number of such closed marginal curves decreases on incrementing $\beta$ from $10^\circ$ to $90^\circ$. The leftmost wide TH marginal curve becomes narrower on increasing $\beta$.
The closed marginal curves do not appear for $\beta$ in the range $10^\circ$--$90^\circ$.
A  small subharmonic (TH) marginal curve appears for $\beta=30^\circ$ to the left of the wider TH branch. The entire pattern of TH, TS, and LH marginal curves shifts upwards in the $(k,\gr)$-plane on incrementing $\beta$ from $0^\circ$ to $90^\circ$, which indicates that in TMILC, the critical Grashof number increases with increasing $\beta$. The critical mode for the onset of TMILC remains  LH for $0^\circ\leq \beta\leq64.806^\circ$, where the angle of inclination $\beta=\beta_c$ denotes the codimension-2 point for the mode switching between longitudinal and transverse types with the following critical values.
 \begin{eqnarray*}
(\beta_c,k_c^{LH},k_c^{TH},\gr_c)=(64.806^\circ,3.0578,2.8475,6693.30).
  \end{eqnarray*}
We provide a  more detailed  description of the marginal curves  for $\beta=\beta_c$ separately in Fig.~\ref{F1b}.
 A meandering pattern of marginal curve is observed in a neighborhood of $k=4.1$. This meandering pattern is found to be a characteristic of the marginal curve not only for $\beta=\beta_c$ but  also for rest of all positive inclinations. Also, the marginal curves for TH or LH responses are found to occur for larger wave numbers, whereas TS response is not observed for large wave numbers.

Coming back to Fig.~\ref{F1a}, we find that for  $\beta\geq \beta_c$, the critical Grashof number always corresponds to TH or TS mode. On further incrementing $\beta$, a bicritical state is observed to occur for  $\beta\approx 71.585^\circ$, where
 \begin{eqnarray*}
(k_c^{TH},k_c^{TS},\gr_c)=(2.9675,2.4675,7025.85).
  \end{eqnarray*}
The instability tongues corresponding to this bicritical state are of comparable sizes. For $\beta=80^\circ$, the critical mode of instability is in the form of a pattern of transverse convective rolls executing subharmonic oscillations.

\begin{figure}[h!!!]
\centering
\includegraphics[width=1\textwidth]{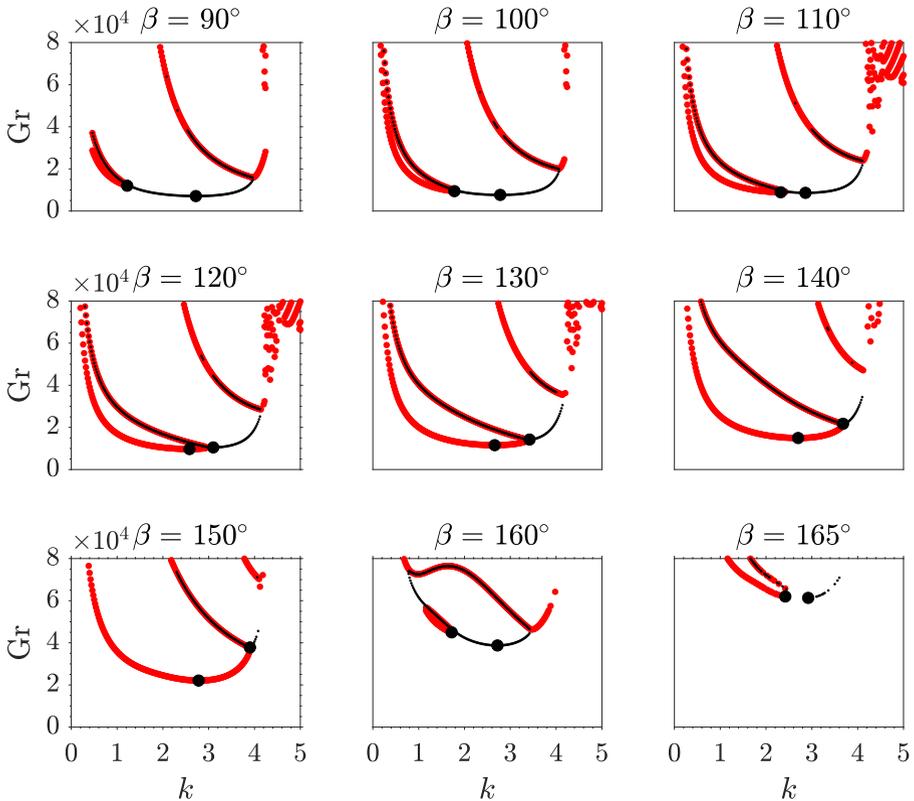}%
 \caption{The neutral instability curves in $(k,\gr)$-plane for the fixed parametric values and the other detail as in Fig.~\ref{F1a}.}\label{F1c}
\end{figure}

The marginal instability curves for $\beta\geq 90^\circ$ have been shown in Fig.~\ref{F1c} for the fixed parametric values as in Fig.~\ref{F1a}.
 For $90^\circ\leq \beta<180^\circ$ the longitudinal mode has not been found to occur in TMILC. Consequently, the instability response is either TH or TS, depending upon $\beta$. The lowest value of Grashof number occurs on the middle subharmonic marginal curve for $\beta=90^\circ,100^\circ,$ and $110^\circ$. In each of these cases, the marginal curves alternate between the middle subharmonic branch accompanied by two left and right harmonic branches. The leftmost harmonic branch is narrow which widens as $\beta$ is incremented, and for $\beta=120^\circ$, $\gr_c$ occurs on the leftmost harmonic marginal curve in the $(k,\gr)$-plane.

\begin{figure}[h!!!]
\centering
\includegraphics[width=\textwidth]{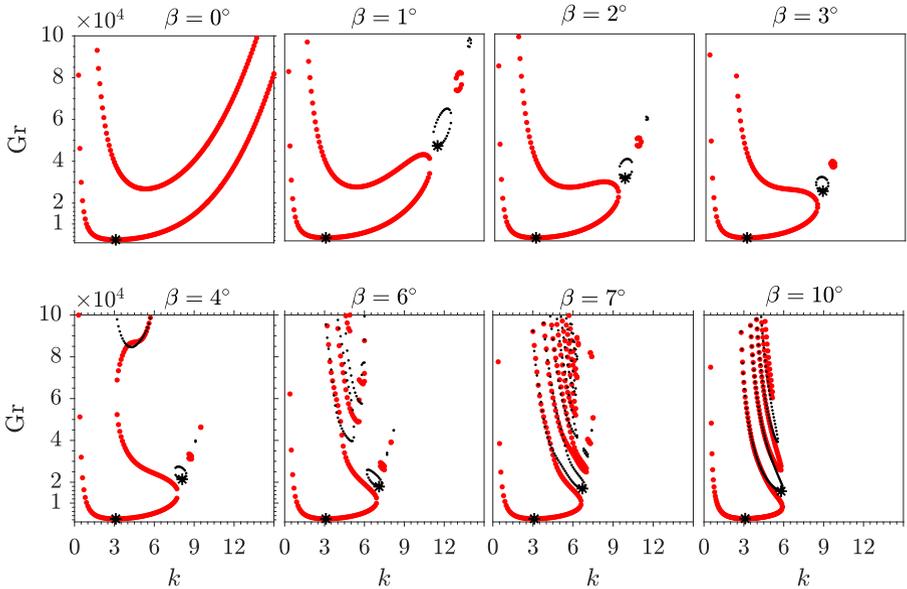}%
 \caption{Marginal curve in $(k,\gr)$-plane for the transverse mode of TMILC for  $\sigma=0.71$, $\epsilon=0.5$, and $\omega=50$. The thicker (red) and thinner (black) parts of the curve correspond to TH and TS responses respectively. Each of the points marked $\bullet$ corresponds to a bicritical state.}\label{F2}
\end{figure}

Under high frequency modulation and small inclinations of the fluid layer, we have observed a beautiful ``Boot-pattern'' of the marginal curves corresponding to the transverse mode. Figure \ref{F2} shows such a pattern  for the Prandtl number of air,  $\epsilon=0.5$ and $\omega=50$.
 For $\beta=0^\circ$, there is a single wide TH instability tongue in the  $(k,\gr)$-plane within which the basic flow is unstable and stable otherwise. For $\beta=1^\circ$, multiple marginal curves alternate between harmonic and subharmonic branches, where there is a leftmost boot-shaped larger harmonic branch on which $Gr_c$ occurs, followed by  alternate subharmonic and harmonic small closed curves. The entire pattern shifts towards the lower wave-numbers on incrementing $\beta$. For $\beta=6^\circ$, a single larger harmonic instability tongue appears along with multiple smaller marginal curves alternating between harmonic and subharmonic types and also a pattern of closed curves alternating between harmonic and subharmonic types. With a further increase of $\beta$, the entire pattern of the smaller instability tongues and the closed loops come closer, and eventually the closed curves disappear for $\beta\approx 10^\circ$.
\subsection{Effect of the angle of inclination on the onset}
\begin{figure}[h!!!]
\centering
\includegraphics[width=\textwidth]{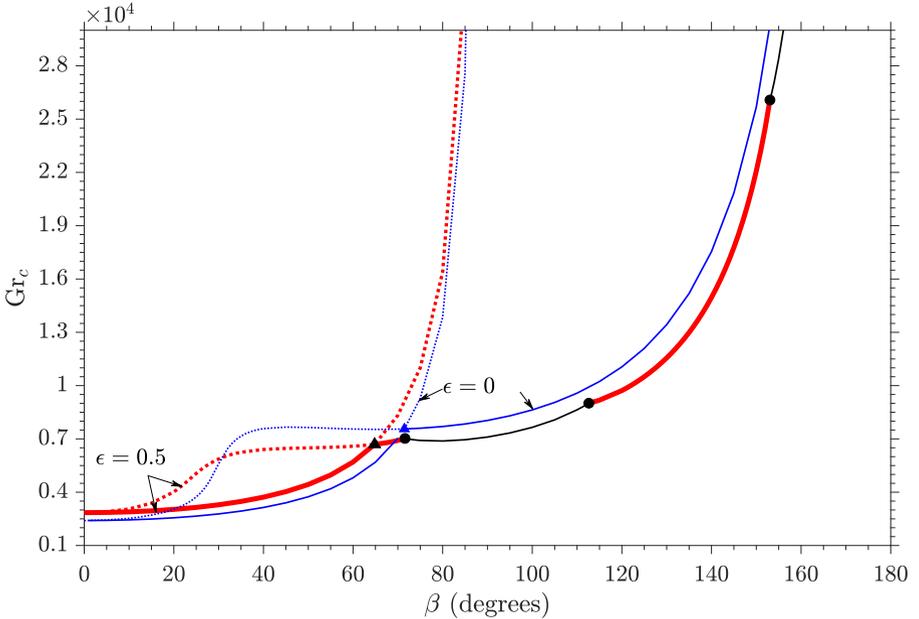}
 \caption{$\gr_c$ vs $\beta$ for $\sigma=0.71$, $\epsilon=0,0.5$, and $\omega=5$. The thinner (magenta) curves and the thicker(red) curves correspond to $\epsilon=0$ and $\epsilon=0.5$, respectively. The thick (blue) curve for $\epsilon=0.5$ corresponds to LH mode. The thicker (red) and thinner (black) points on the curve $\epsilon=0.5$ correspond to TS and TH   modes, respectively.  The points marked $\bullet$ correspond to  bicritical states. The points marked $\blacktriangle$  on curves $\epsilon=0$ and $\epsilon=0.5$ correspond to $\beta_c$.}\label{F3a}
\end{figure}

Figure \ref{F3a} shows the variation of $\gr_c$ with $\beta$ for $\sigma=0.71$, $\epsilon=0,0.5$, and $\omega=5$.
 For the unmodulated ILC, that is for $\epsilon=0$, the instability response is a longitudinal mode for $\beta$ in the range $0^\circ$--$71.44^\circ$, whereas the instability response is a transverse mode for $71.44^\circ<\beta<180^\circ$. Also, $\gr_c$ is an increasing function of $\beta$ where the increase is   abrupt in a neighborhood of $\beta=180^\circ$.

A similar variation of $\gr_c$ with $\beta$ is found to occur  under modulation  for $\epsilon=0.5$ and $\omega=5$. Here, the preferred mode for the onset of TMILC is a LH mode for $0^\circ\leq \beta<64.806^\circ$. The angle of inclination $64.806^\circ$ corresponds to the codimension-2 point (See Fig.~\ref{F1b}). For $64.806^\circ<\beta<71.585^\circ$, the instability response is a transverse harmonic mode, and a bicritical state is observed to occur for $\beta\approx 71.585^\circ$ with the following critical values.
\begin{eqnarray*}
(k_c^{TH},k_c^{TS},\gr_c)=(2.9675,2.4675,7025.846).
\end{eqnarray*}
For $71.585^\circ<\beta<112.63^\circ$, the instability response in TMILC is a TS mode till another bicritical state occurs for
\begin{eqnarray*}
(\beta,k_c^{TS},k_c^{TH},\gr_c)=(112.63^\circ,2.9143,2.4300,9011.171).
\end{eqnarray*}
For $112.63^\circ<\beta<153.0470^\circ$, the instability response is a TH mode. Yet another bicritical state is observed for  $\beta\approx153.0470^\circ$ with
\begin{eqnarray*}
(k_c^{TH},k_c^{TS},\gr_c)=(2.9060, 2.4690, 26076.20).
\end{eqnarray*}
Beyond $\beta=153.0470^\circ$, the instability response is found to be a TS mode. It is interesting to observe that
\begin{eqnarray*}
\gr_c\mid_{\epsilon=0}
\begin{cases}&<\gr_c\mid_{\epsilon=0.5} ~\text{for}~0^\circ\leq \beta<70^\circ;\\
&>\gr_c\mid_{\epsilon=0.5} ~\text{for}~70^\circ<\beta<180^\circ.
\end{cases}
\end{eqnarray*}
This shows that the onset of TMILC gets delayed by modulation for $0^\circ\leq \beta<71.585^\circ$ and the onset gets advanced under modulation for $71.585^\circ<\beta<180^\circ$.

\begin{figure}[h!!!]
\centering
\includegraphics[width=\textwidth]{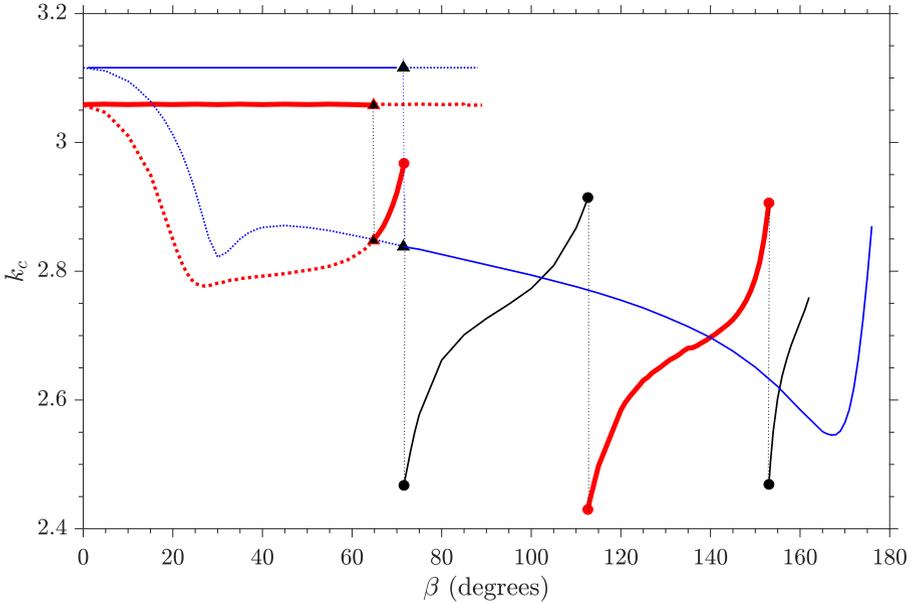}
 \caption{$k_c$ vs $\beta$ for the fixed parametric values and other detail as in Fig.~\ref{F3a}.}\label{F3b}
\end{figure}

The corresponding variation of $k_c$ with $\beta$ is shown in Fig.~\ref{F3b}.
The critical wave number $k_c^{L}$ or $k_c^{LH}$ for the onset of TMILC is independent of $\beta$ under the absence as well as the presence of the modulation. However,  the critical wave number corresponding to the transverse mode is a nonconstant function of $\beta$. For $\epsilon=0$, $k_c^T$ is a decreasing continuous function of $\beta$ upto $168^\circ$ approximately, beyond which $k_c^T$ increases sharply with $\beta$ on further increase of $\beta$. Under modulation, a discontinuity occurs in $k_c$ for the value of $\beta$  at which a bicritical state occurs in TMILC. Such a bicritical state is responsible for a mode switching between harmonic and subharmonic types. Between any two consecutive bicritical states, $k_c$ is an increasing function of $\beta$.
\subsection{Effect of modulation on mode switching}
\begin{figure}[h!!!]
\centering
\includegraphics[width=\textwidth]{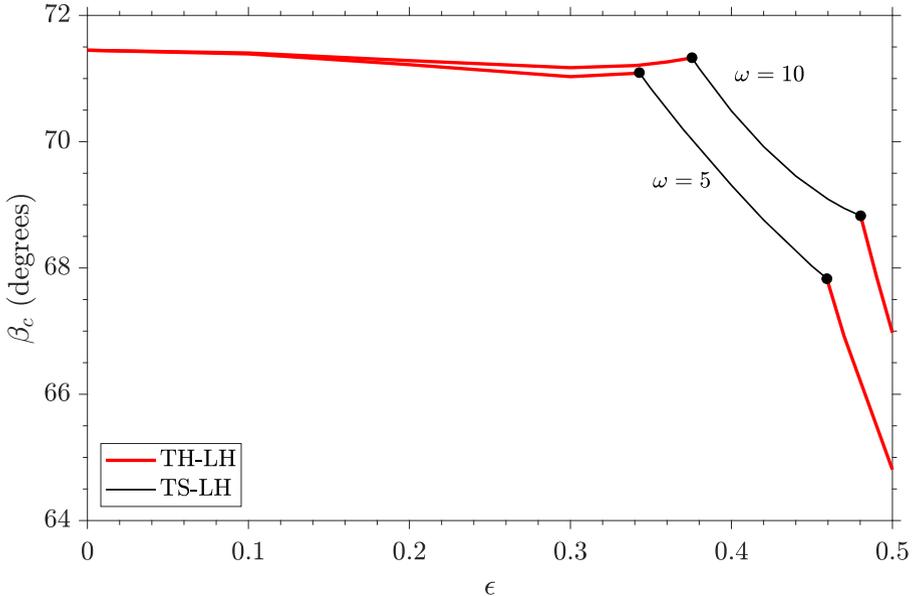}%
 \caption{$\beta_c$ vs $\epsilon$ for $\sigma=0.71$ and $\omega=5,10$. The thicker (red) and thinner (black) parts of the curve correspond to harmonic and subharmonic responses respectively. Each of the points marked $\bullet$ corresponds to a bicritical state.}\label{F4}
\end{figure}

The angle of inclination for the codimension-2 point for the steady ILC is $\beta_c\approx 71.4483^\circ$ for the Prandtl number of air, with the following critical values
\begin{eqnarray*}
(k_c^{L},k_c^T,\gr_c)\approx (3.116,2.838,7560.021990).
\end{eqnarray*}
Under modulation, the parameter $\beta_c$ is a function of the modulation parameters. In this view, we have obtained Fig.~\ref{F4}, which shows the variation of $\beta_c$  with $\epsilon$ for  $\sigma=0.71$, and for $\omega=5,10$. The curve for each value of $\omega$ consists of alternate harmonic and subharmonic parts separated by an intermediate bicritical state.

First we explain the curve $\omega=5$. The parameter $\beta_c$ decreases slightly from $71.4483^\circ$ to $71.0300^\circ$, when $\epsilon$ is varied from $0$ to $0.3$ approximately, where the instability response of TMILC is harmonic. For $0.3\leq \epsilon\leq 0.3428$, the parameter $\beta_c$ increases slightly with $\epsilon$, and a bicritical state is observed to occur for $\epsilon\approx 0.3428$ with the following critical values.
\begin{eqnarray*}
(\beta_c,k_c^{TH},k_c^{TS},k_c^{LH},\gr_c)\approx (71.094^\circ,3.019,2.579,3.104,7903.3).
\end{eqnarray*}
For $0.3428<\epsilon<0.4593$, the instability response is subharmonic and $\beta_c$ decreases significantly to $67.830^\circ$ for $\epsilon=0.4528$ when another bicritical state occurs for the following critical values
\begin{eqnarray*}
(\epsilon,\beta_c,k_c^{TS},k_c^{TH},k_c^{LH},\gr_c)\approx (0.4528,67.83^\circ,2.486,2.980,3.081,7264.6).
\end{eqnarray*}
Beyond $\epsilon=0.4593$, $\beta_c$ decreases rapidly with $\epsilon$ and the instability response remains harmonic. These observations show that under modulation, the inclination corresponding to the codimension-2 point is either harmonic, or subharmonic, or bicritical, depending upon the amplitude of modulation.

A similar variation of $\beta_c$ with $\epsilon$ occurs for $\omega=10$. Here, the instability region for the onset of longitudinal mode expands in comparison with the one that corresponds to $\omega=5$.

\begin{figure}[h!h!h!]
\centering
\includegraphics[width=\textwidth]{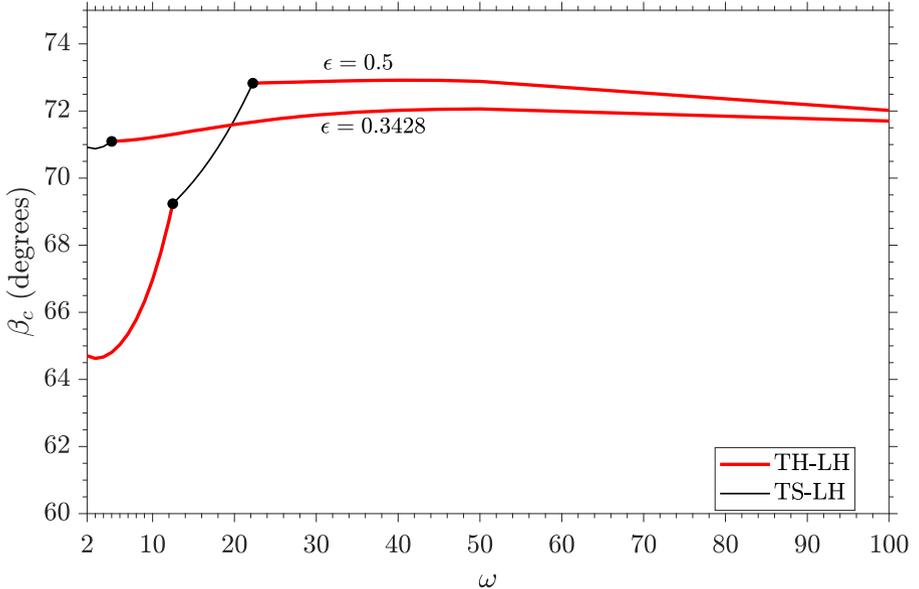}%
 \caption{$\beta_c$ vs $\omega$ for $\sigma=0.71$ and $\epsilon=0.5, 0.3428$. The thicker (red) and thinner (black) parts of the curve correspond to TH-LH and TS-LH responses respectively. Each of the points marked $\bullet$ corresponds to a bicritical state.}\label{F5}
\end{figure}

To see the dependence of $\beta_c$ on $\omega$, we have obtained Fig.~\ref{F5} for two distinct values of $\epsilon=0.3428,0.5$ and $\sigma=0.71$. The thicker (red) and thinner (black) parts of each of the curves in Fig.~\ref{F5} correspond to harmonic and subharmonic responses respectively. Each of the points marked $\bullet$ corresponds to a bicritical state. It is evident from Fig.~\ref{F5} that the locus of the angle of inclination corresponding to the codimension-2 point in the $(\omega,\gr_c)$-plane for TMILC consists of alternating TH-LH and TS-LH branches connected through an intermediate bicritical state.  First we  explain the curve $\epsilon=0.3428$. For $2\leq \omega<5$, the codimension-2 point occurs through a TS-LH response, where $\beta_c$ decreases with $\omega$ for $2\leq \omega<4$, attains a minimum for $\omega\approx 4$, and increases with omega for $4<\omega<5$. For $\omega=5$ the codimension-2 point corresponds to a bicritical state as discussed earlier for $\beta_c\approx 71.094^\circ$. For $\omega>5$, the value of $\beta_c$ corresponds to a TH-LH response, where $\beta_c$ increases with $\omega$ for $5<\omega<50$ and attains a maximum value $\beta_c\approx 72.0620^\circ$ for $\omega\approx 50$. Only $1.36\%$ increase in $\beta_c$ with $\omega$ is observed to occur for $\omega>5$, which shows that in TMILC, the high frequency modulation will have little effect on the angle of inclination corresponding to the codimension-2 point. Beyond $\omega=50$, $\beta_c$ decreases slowly with a further increase of $\omega$ and eventually approaches the following:
\begin{eqnarray*}
\lim_{\omega\rightarrow \infty} \beta_c(\omega)&=& 71.4483^\circ=\lim_{\epsilon\rightarrow 0}\beta_c.
\end{eqnarray*}
A similar dependence of $\beta_c$ on $\omega$ can be observed for $\epsilon=0.5$, where the dependence is  significant for $2\leq \omega\leq22.254$. Here, $\beta_c$ corresponds to a TH-LH response for $2\leq \omega< 12.469$, and a bicritical state is observed for
\begin{eqnarray*}
(\omega,\beta_c,k_c^{TH},k_c^{TS},k_c^{LH},\gr_c)&=&(12.469,69.237^\circ,3.018,2.239,3.071,7715.8).
\end{eqnarray*}
The parameter $\beta_c$ decreases with $\omega$ for $2\leq \omega\leq 3$, attains a minimum for $\omega=3$, and then $\beta_c$  increases rapidly with $\omega$ for $3\leq \omega\leq 22.254$. For $12.469<\omega<22.254$, $\beta_c$ corresponds to a TS-LH response, when another bicritical state occurs for $\omega\approx 22.254$, which corresponds to the following values
\begin{eqnarray*}
(\beta_c,k_c^{TS},k_c^{TH},k_c^{LH},\gr_c)&=&(72.829^\circ,2.023,3.009,3.081,8837.8).
\end{eqnarray*}
Beyond $\omega=22.254$, the variation in $\beta_c$ with $\omega$ for $\epsilon=0.5$ is similar as in the case of $\epsilon=0.3428$.
\section{Prandtl number dependence}\label{sec:4}
\begin{figure}[h!!!]
\centering
\includegraphics[width=\textwidth]{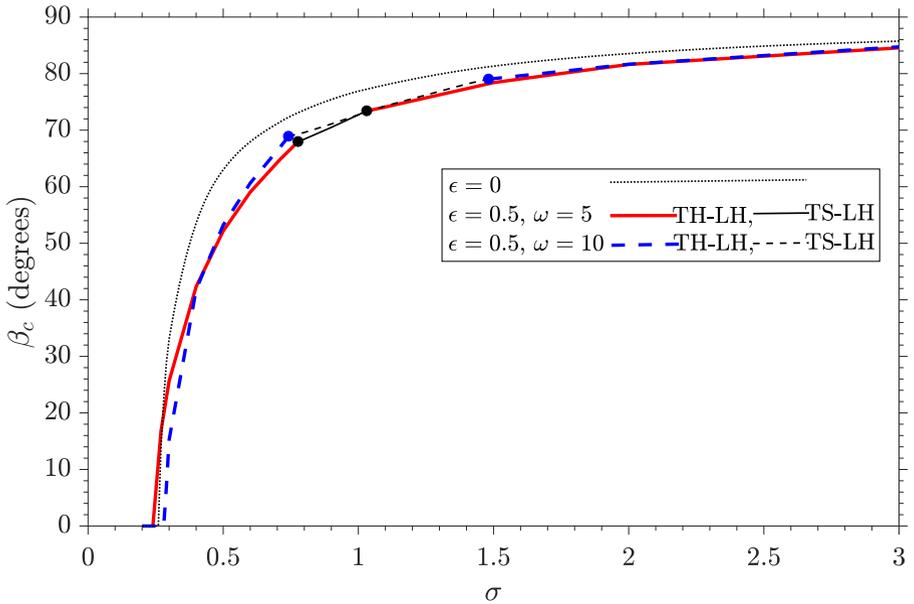}%
 \caption{$\beta_c$ vs $\sigma$ for $\epsilon=0,0.5$, and $\omega=5,10$. The dotted curve corresponds to $\epsilon=0$. The  thicker (red) and  thinner (black) solid curves are drawn for  $(\epsilon,\omega)=(0.5,5)$. The  thicker (blue) and  thinner (black) dashed curves are drawn for  $(\epsilon,\omega)=(0.5,10)$. The thicker and thinner curves for $\epsilon=0.5$ correspond to for TH-LH and TS-LH responses respectively. Each of the points marked $\bullet$ corresponds to a bicritical state.}\label{F6}
\end{figure}

Figure \ref{F6} shows variation of $\beta_c$ with $\sigma$ for $\epsilon=0,0.5$ and $\omega=5, 10$. The dotted, dashed, and solid curves correspond to $\epsilon=0$, $(\epsilon,\omega)=(0.5, 5)$, and $(\epsilon,\omega)=(0.5, 10)$, respectively. The thicker (red, blue) and thinner (black) parts of each of the curves for $\epsilon=0.5$ correspond to TH-LH and TS-LH responses, respectively. In the absence of modulation ($\epsilon=0$), the parameter $\beta_c$ is a monotonically increasing function of $\sigma$, where the rate of increase of $\beta_c$ with respect to $\sigma$ is sharp for small values of $\sigma$ and the rate is negligible for large values of $\sigma$. It is well known that for the unmodulated ILC, $\beta_c=0^\circ$ for all fluids with $\sigma<0.26$ and $\beta_c>0$ for $\sigma\geq 0.26$  approximately \cite{korpela1974,chen1989}.

Under modulation ($\epsilon=0.5$), a similar observation prevails. In fact the variation of $\beta_c$ with $\sigma$ is similar in the two typical cases $\epsilon=0$ and $\epsilon=0.5$ except that for $\epsilon=0.5$, the nature of the onset of the instability for $\beta_c$ can be one of the three modes TH-LH, TS-LH, and TH-TS-LH. More precisely, for $\epsilon=0.5$ and $\omega=5$, $\beta_c>0^\circ$ for all $\sigma\geq 0.24$ and $\beta_c=0^\circ$ otherwise. Also, $\beta_c$ corresponds to TH-LH mode for $0<\sigma<0.7766$ and a bicritical state (a TH-TS-LH mode) with
\begin{eqnarray*}
(\sigma,\beta_c,k_c^{TH},k_c^{TS},k_c^{LH},\gr_c)&\approx&(0.7766,67.996^\circ,2.971,2.492,3.051,6921.34).
\end{eqnarray*}
For $0.7766<\sigma<1.0308$, the parameter $\beta_c$ corresponds to a TS-LH mode. The value $\sigma=1.0308$ corresponds to another bicritical state with the following numerical values
\begin{eqnarray*}
(\beta_c,k_c^{TS},k_c^{TH},k_c^{LH},\gr_c)&\approx&(73.4314^\circ,2.900,2.478,3.038,6719.5).
\end{eqnarray*}
For $\sigma>1.0308$, $\beta_c$ corresponds to a TH-LH mode where for all sufficiently large values of $\sigma$, the rate of change of $\beta_c$ with $\sigma$ is negligible.

A similar variation of $\beta_c$ with $\sigma$ occurs for $\epsilon=0.5$ and $\omega=10$ (dashed curves), where we note that $\beta_c=0^\circ$ for $\sigma<0.279$ and $\beta_c>0^\circ$ otherwise. Also, $\beta_c$ corresponds to a TH-LH mode for $0<\sigma<0.7414$, a TS-LH mode for  $0.7414<\sigma<1.4815$, and  a TH-TS-LH mode for $\sigma=0.7414, 1.4815$. The curves $\omega=5$ and $\omega=10$ are practically indistinguishable for $\sigma>1$, which indicates that under modulation, an increase of $\omega$ has negligible effect on $\beta_c$  for the fluids having large Prandtl number.
\section{Concluding remarks}\label{sec:5}
The Floquet analysis of TMILC reveals interesting pattern of the nature of the onset of the instability.

The marginal instability curve for the longitudinal perturbations in TMILC consists of a single wide instability tongue in the space of the wave number of the perturbations and the Grashof number, and the instability response is harmonic.

On the other hand, the marginal instability curves for the transverse perturbations in TMILC consist of an alternate pattern of harmonic and subharmonic regions in the space of the wave number of the perturbations and the Grashof number. In this case the instability response of TMILC can be either harmonic, or subharmonic, or bicritical, depending upon the modulation parameters and the angle of inclination of the layer with respect to the horizontal.

The onset mode of TMILC is found to be a time periodically oscillating pattern of convective rolls which corresponds to any of the following six modes:  TH, TS, LH, TH-LH, TS-LH, TH-TS-LH, depending upon the orientation of the layer with respect to the horizontal and the modulation parameters.  The modulation effects are significant for low to moderate forcing frequencies, and are negligible for sufficiently large frequencies.

Except for a neighborhood of the angle of inclination for the codimension-2 point, the modulation delays the onset of the longitudinal mode of TMILC and the modulation favors the onset of the transverse mode of TMILC.

The angle of inclination for the codimension-2 point is also a function of the modulation parameters and and Prandtl number of the fluid.
The rate of variation of the angle of inclination for the codimension-2 point  with the Prandtl number is significant for the low Prandtl number fluids.

The present analysis on TMILC is  mainly focussed on investigating the effect of modulation on the value of the angle of inclination of the fluid layer with respect to the horizontal below which the preferred onset mode of TMILC is longitudinal and above which the preferred onset mode is transverse.  For such an inclination, Floquet theory allows us to further classify the nature of the onset of the instability in the form of a time-periodic flow among the harmonic and the subharmonic types.

However, for large Prandtl number fluids, the onset of the instability even in the unmodulated ILC can be an oscillatory mode for $\sigma>12.5$. Under modulation, such a mode becomes a quasi-periodic mode. Consequently, for a fluid with $\sigma>12.5$, a further investigation of the onset of TMILC  is needed to be done to include the possibility of a quasi-periodic mode, and we leave this task for future research.

\backmatter
%\bmhead{Not Applicable}
\bmhead{Acknowledgments}
The present research work is supported by Science and Engineering Research Board (SERB), Government of India under MATRICS Scheme wide project grant no. \textbf{MTR/2017/000575} awarded to the author.
\section*{Declarations}
\noindent
The author declares that he has no competing interest.

\begin{appendices}

\section{}\label{appen1}



To define various matrices and block matrices, we proceed as follows:
Let $O$ and $I$  respectively denote the zero matrix and the identity matrix each of order $N\times N$. Also, for $X_\ell,~Y_j\in \mathcal{L}^2(-\frac{1}{2},\frac{1}{2})$ with $1\leq \ell ,j\leq N$, we define $\bigl<X_\ell,Y_j\bigr>$ as the $N\times N$ matrix whose $(\ell,j)$th entry is equal to
%\begin{equation*}
    $\int_{-\frac{1}{2}}^{\frac{1}{2}}X_\ell(x)\overline{Y_j(x)}dx$. Having said this,
the following  block matrices have been used in \eqref{e170} for each $q=-L,\ldots,L$.
\begin{eqnarray}\label{eq18}
    \mathbf{L}_q=\frac{\iota\omega(s+q)}{\sigma}\left(
                   \begin{array}{cccc}
                     L_1(q)-A_1 &O&O&O \\
                     O&L_2(q)-B_2&O&O\\
                     O&-\bigl<\Psi_\ell,\si_j\bigr>&\frac{\sigma}{2}I-C_3&O\\
                     -\langle\Phi_\ell,\cs_j\rangle&O&O&\frac{\sigma}{2}I-D_4\\
                   \end{array}
                 \right)-\mathbf{U}_0,
\end{eqnarray}
%\begin{eqnarray}
%    \mathbf{U}_0=\left(
%                   \begin{array}{cccc}
%                     A_1 &O&O&O \\
%                     O&B_2&O&O\\
%                     O&\bigl<\Psi_\ell,\si_j\bigr>&C_3&O\\
%                     \bigl<\Phi_\ell,\cs_j\bigr>&O&O& D_4\\
%                   \end{array}
%                 \right),
%\end{eqnarray}
\begin{eqnarray}
    \mathbf{U}_1=k^2\left(
                   \begin{array}{cccc}
                     O &O&O&\langle\cs_\ell,\Phi_j\rangle \\
                     O&O&\langle\si_\ell,\Psi_j\rangle&O\\
                     O&O&O&O\\
                     O&O&O&O\\
                   \end{array}
                 \right),
\end{eqnarray}
\begin{eqnarray}
 \mathbf{U}_2=\frac{k}{6}\left(
                   \begin{array}{cccc}
                     O &A_2&6A_3&O \\
                     B_1&O&O&6B_4\\
                     O&O&O&-\bigl<h(x)\cs_\ell,\si_j\bigr>\\
                     O&O&-\bigl<h(x)\si_\ell,\cs_j\bigr>&O\\
                   \end{array}
                 \right),
\end{eqnarray}
\begin{eqnarray}\label{eq190}
    \mathbf{V}=\left(
                   \begin{array}{cccc}
                     O &O&O&O \\
                     O&O&O&O\\
                     O&G_{2}&O&O\\
                     H_1&O&O&O\\
                   \end{array}
                 \right);
%\end{eqnarray}
%\begin{eqnarray}
\label{eq19}
    \mathbf{W}=\frac{k}{2\iota\omega(1-\sigma)}\left(
                   \begin{array}{cccc}
                     O &E_2&O&O \\
                     F_1&O&O&O\\
                     O&O&O& \sigma G_4\\
                     O&O&\sigma H_3&O\\
                   \end{array}
                 \right), \sigma\neq 1.
\end{eqnarray}
%\end{equation*}
Letting  $h(x)=x\bigl(x^2-\frac{1}{4}\bigr)$, we have for all $1\leq j, \ell\leq N$ the following matrices.
\begin{eqnarray}
    \label{eq22} L_{1}(q)&=&\bigl<\Phi_\ell'',\Phi_j\bigr>-k^2\langle\delta_{\ell j}\rangle;~ L_{2}(q)=\bigl<\Psi_\ell'',\Psi_j\bigr>-k^2\langle\delta_{\ell j}\rangle,\\
      \label{eq25}   A_1&=&\langle(\lambda_\ell^4+k^4)\delta_{\ell j}\rangle-2k^2\langle\Phi_\ell'',\Phi_j\rangle,\\
   \label{eq26}A_2 &=&\frac{1}{\sigma}\bigl\{\bigl<h(x)\Psi_\ell'',\Phi_j\bigr>-\bigl<h''(x)\Psi_\ell,\Phi_j\bigr> -k^2 \bigl<h(x)\Psi_\ell,\Phi_j\bigr>\bigr\},\\
   \label{eq27}A_3 &=&2\pi \bigl<\ell\cos \{2\ell\pi x\},\Phi_j\bigr>,\\
%   \end{eqnarray}
%\begin{eqnarray}
   \label{eq28}B_1 &=&\frac{1}{\sigma}\bigl\{\bigl<h''(x)\Phi_\ell,\Psi_j\bigr>-\bigl<h(x)\Phi_\ell'',\Psi_j\bigr>+k^2 \bigl<h(x)\Phi_\ell,\Psi_j\bigr>\bigr\},\\
\label{eq31}
    B_2 &=&\langle(\mu_\ell^4+k^4)\delta_{\ell j}\rangle-2k^2\bigl<\Psi_\ell'',\Psi_j\bigr>,\\
    \label{eq32}B_{4} &=&\pi \bigl<(2\ell-1)\sin\{(2\ell-1)\pi x\},\Psi_j\bigr>,\\
      C_{3} &=&-\frac{1}{2}\langle(4\pi^2\ell^2+k^2)\delta_{\ell j}\rangle;~
      D_{4} =-\frac{1}{2}\langle((2\ell-1)^2\pi^2+k^2)\delta_{\ell j}\rangle.
\end{eqnarray}
For $\sigma>0$ and $X_\ell$, $Y_j$ in $\mathcal{L}^2(-\frac{1}{2},\frac{1}{2})$, we define $\langle M_\sigma(X_\ell,Y_j)\rangle$ and $\langle N_\sigma(X_\ell,Y_j)\rangle$
as the $N\times N$ matrices whose $(\ell,j)$th entries are
 $\int_{-1/2}^{1/2}f_\sigma(x,\omega) X(x)\overline{Y(x)}dx$ and
    $\int_{-1/2}^{1/2}X(x)\frac{\partial}{\partial x}f_\sigma(x,\omega)\overline{Y(x)}dx$, respectively.
The following matrix entries are used in \eqref{eq190}.
\begin{eqnarray}\label{eq40}
    E_{2} &=&\langle M_1(\Psi_\ell'',\Phi_j)\rangle-(\iota\omega+k^2)\langle M_1(\Psi_\ell,\Phi_j)\rangle\\
    \nonumber&&+\Bigl(\frac{\iota\omega}{\sigma}+k^2\Bigr)\langle M_\sigma(\Psi_\ell,\Phi_j)\rangle
    -\langle M_\sigma(\Psi_\ell'',\Phi_j)\rangle,\\
\label{eq43}
   F_{1} &=&(\iota\omega+k^2)\langle M_1(\Phi_\ell,\Psi_j)\rangle
   -\langle M_1(\Phi_\ell'',\Psi_j)\rangle\\
   \nonumber &&-\Bigl(\frac{\iota\omega}{\sigma}+k^2\Bigr)\langle M_\sigma(\Phi_\ell,\Psi_j)\rangle
   +\langle M_\sigma(\Phi_\ell'',\Psi_j)\rangle,\\
   G_{2} &=&-\frac{1}{2}\langle N_1(\Psi_\ell,\si_j)\rangle;~
  G_{4} =-\langle M_1(\cs_\ell,\si_j)\rangle+\langle M_\sigma(\cs_\ell,\si_j)\rangle,\\
    H_{1} &=&-\frac{1}{2}\langle N_1(\Phi_\ell,\cs_j)\rangle;~
    H_{3} =\langle M_1(\si_\ell,\cs_j)\rangle-\langle M_\sigma(\si_\ell,\cs_j)\rangle.
  \end{eqnarray}
\end{appendices}


%\bibliography{references2}% common bib file
%\input refernces2.bbl


\end{document}